\documentclass[pra,twocolumn,superscriptaddress,floatfix]{revtex4-1}
\usepackage{amsmath,amsfonts,amssymb,amsthm,graphics,graphicx,epsfig,bbm}
\usepackage[colorlinks=true,citecolor=blue,linkcolor=blue,urlcolor=blue]{hyperref}
\usepackage[usenames]{color}
\usepackage{graphicx}
\usepackage{subfigure}
\usepackage{amsmath}
\usepackage{epsfig}
\usepackage{dcolumn}
\usepackage{bm}
\usepackage{color}
\usepackage{verbatim}
\usepackage{epstopdf}
\usepackage{amssymb} 
\usepackage{amstext}
\usepackage{latexsym}
\usepackage{hyperref}
\usepackage{amsfonts}
\usepackage{psfrag}
\usepackage{xcolor}
\usepackage{times}
\usepackage{dsfont }

\newcommand{\ket}[1]{\ensuremath{\left|{#1}\right\rangle}}
\newcommand{\bra}[1]{\ensuremath{\left\langle{#1}\right|}}

\begin{document}
\title{Parity-dependent State Engineering and Tomography in the ultrastrong coupling regime}
\author{S. Felicetti}
\affiliation{Department of Physical Chemistry, University of the Basque Country UPV/EHU, Apartado 644, E-48080 Bilbao, Spain}
\author{T. Douce}
\affiliation{Laboratoire Mat\'eriaux et Ph\'enom\`enes Quantiques, Universit\'e Paris Diderot, CNRS UMR 7162, 75013, Paris, France}
\author{G. Romero}
\affiliation{Department of Physical Chemistry, University of the Basque Country UPV/EHU, Apartado 644, E-48080 Bilbao, Spain}
\affiliation{Departamento de F\'isica, Universidad de Santiago de Chile (USACH), Avenida Ecuador 3493, 917-0124, Santiago, Chile}
\author{P.  Milman}
\affiliation{Laboratoire Mat\'eriaux et Ph\'enom\`enes Quantiques, Universit\'e Paris Diderot, CNRS UMR 7162, 75013, Paris, France}
\author{E. Solano}
\affiliation{Department of Physical Chemistry, University of the Basque Country UPV/EHU, Apartado 644, E-48080 Bilbao, Spain}
\affiliation{IKERBASQUE, Basque Foundation for Science, Maria Diaz de Haro 3, 48013 Bilbao, Spain}

\date{\today}

\begin{abstract} 
Reaching the strong coupling regime of light-matter interaction has led to an impressive development in fundamental quantum physics and applications to quantum information processing. Latests advances in different quantum technologies, like superconducting circuits or semiconductor quantum wells, show that the ultrastrong coupling regime (USC) can also be achieved, where novel physical phenomena and potential computational benefits have been predicted. Nevertheless, the lack of effective decoupling mechanism in this regime has so far hindered control and measurement processes. Here, we propose a method based on parity symmetry conservation that allows for the generation and reconstruction of arbitrary states in the ultrastrong coupling regime of light-matter interactions. Our protocol requires minimal external resources by making use of the coupling between the USC system and an ancillary two-level quantum system. 
\end{abstract}

\date{\today}

\maketitle

\section{Introduction}
The realization of platforms composed of effective two-level quantum systems interacting with the discrete electromagnetic modes of a resonator represents a milestone in the history of quantum physics. In particular, the achievement of the strong coupling (SC) regime, in which light-matter coupling overcomes losses, gave birth to the field of cavity quantum electrodynamics~\cite{Walther2006,QuantumBook}. Recent experimental developments have shown that the ultrastrong coupling (USC) regime, a limit of the quantum Rabi model (QRM)~\cite{Rabi36,Braak2011}, can also be achieved in a number of implementations such as superconducting circuits~\cite{Bourassa2009, Niemczyk2010, Fedorov2010, Diaz2010}, semiconductor quantum wells~\cite{Gunter2009, Todorov2010, Anappara2009}, and possibly in surface acoustic waves~\cite{Gustafsson2014}. The USC regime is characterized by a coupling strength between the cavity field and matter qubits which is comparable with the resonator frequency. In this case, the field and the two-level system merge into collective bound states, called polaritons. Among other features, the aforementioned polaritons exhibit multiphoton entangled ground states~\cite{Ciuti2005} and parity protection~\cite{Nataf2011}. These represent the distinctive behavior of the USC regime when compared with the SC regime.

The fast-growing interest in the USC regime is motivated by theoretical predictions of novel fundamental properties~\cite{Emary2004, Ciuti2005, Ciuti2006, Deliberato2009, Meaney2010, Ashhab2010, Ridolfo2012,Felicetti2014}, and by potential applications in quantum computing tasks~\cite{Nataf2011, Romero2012, Kyaw2014}. Nowadays, quantum technologies featuring the USC regime have been able to characterize this coupling regime by means of transmission or reflection spectroscopy measurements of optical/microwave signals~\cite{Niemczyk2010,Diaz2010}. However, state reconstruction in the USC regime of the QRM, as well as quantum information applications, are hindered by the lack of \textit{in situ} switchability and control of the cavity-qubit coupling strength. Direct Wigner function reconstruction of an anharmonic oscillator has been realized~\cite{Shalibo2013}, but only for a small anharmonicity. In the case of harmonic oscillators, microwave cavity field states have been measured using streaming Rydberg atoms as probe~\cite{Bertet2002, Deleglise2008}. 

\begin{figure}[]
\centering
\includegraphics[width=0.8\columnwidth]{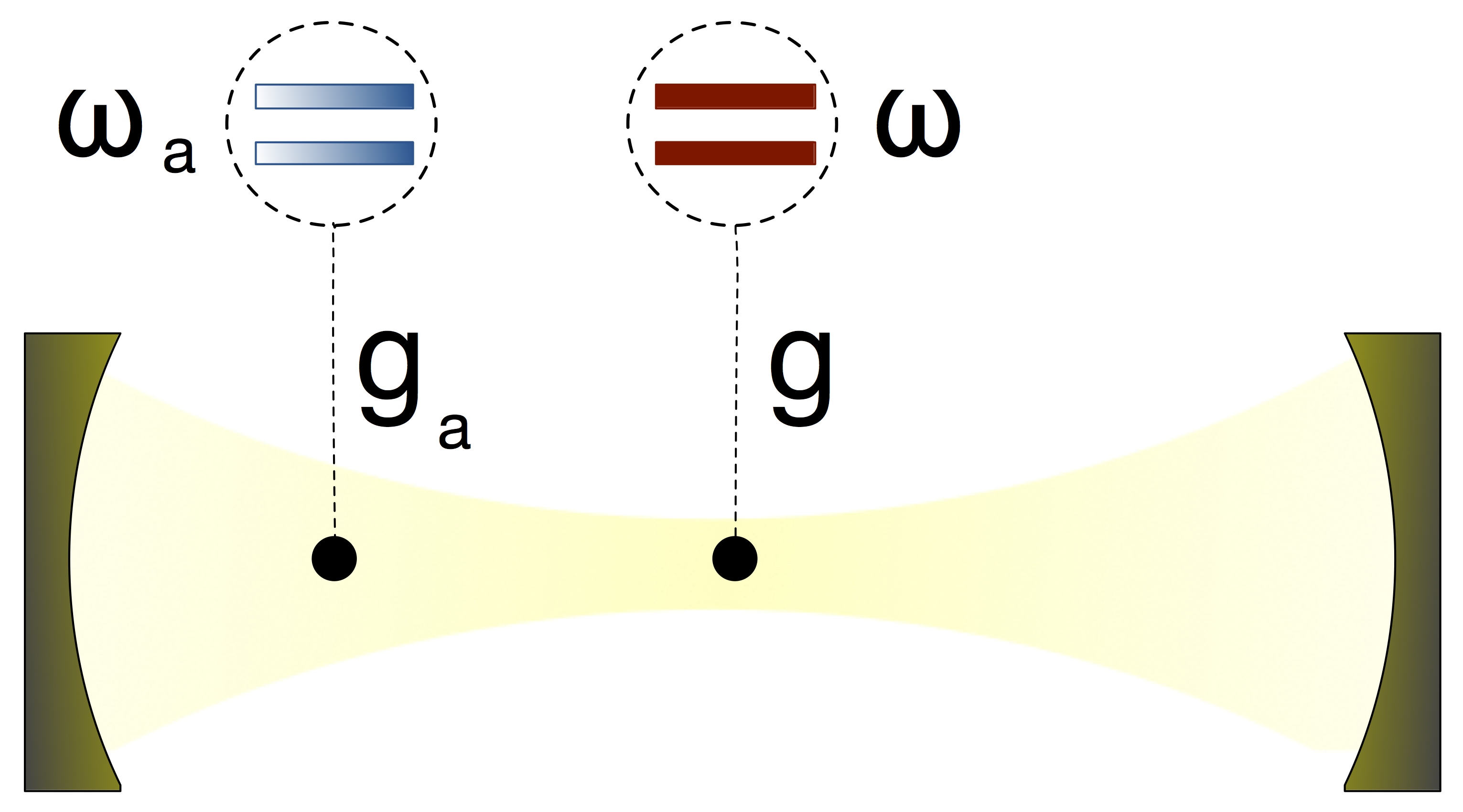}
\caption{{\bf Sketch of a hypothetical quantum-optical implementation of the proposed system.}  A single-mode quantum optical cavity interacts with a qubit (red, solid color) of frequency $\omega$ in the ultrastrong coupling regime. The coupling $g$ is of the same order of the qubit and resonator  frequencies. Another qubit (blue, shaded color) can be used as an ancillary system in order to measure and manipulate USC polariton states.}
\label{sketch}
\end{figure}

Here, we propose the use of an ancillary qubit as a tool for state generation, spectroscopy, and state reconstruction of USC polariton states. We analyze a system composed of a single-mode quantum resonator coupled to two two-level systems, or qubits, as shown in Fig.~\ref{sketch}. One of them (system qubit) interacts with the cavity mode in the USC regime, forming polariton states, while the coupling strength of the ancillary qubit with the cavity is in the SC regime.  Our analysis enables us to design a spectroscopy protocol able to identify the parity of each USC energy level, allowing us to check distinctive features of the USC spectrum in a realistic experiment. Moreover, we show how the ancillary qubit allows for state engineering and tomography of the USC qubit-cavity system. From our analysis, it emerges that USC polaritons populating the system substantially modify the light-matter interaction of the ancillary qubit, leading to a counter-intuitive breaking of the Jaynes-Cummings model~\cite{JCmodel} even for small interaction strengths. Finally, we consider realistic parameters of current implementations of circuit QED in the USC regime, where the present model may be implemented with state-of-the-art technology.

\section{ The quantum Rabi model and an ancillary qubit}
\begin{figure}[]
\centering
\includegraphics[width=0.7\columnwidth]{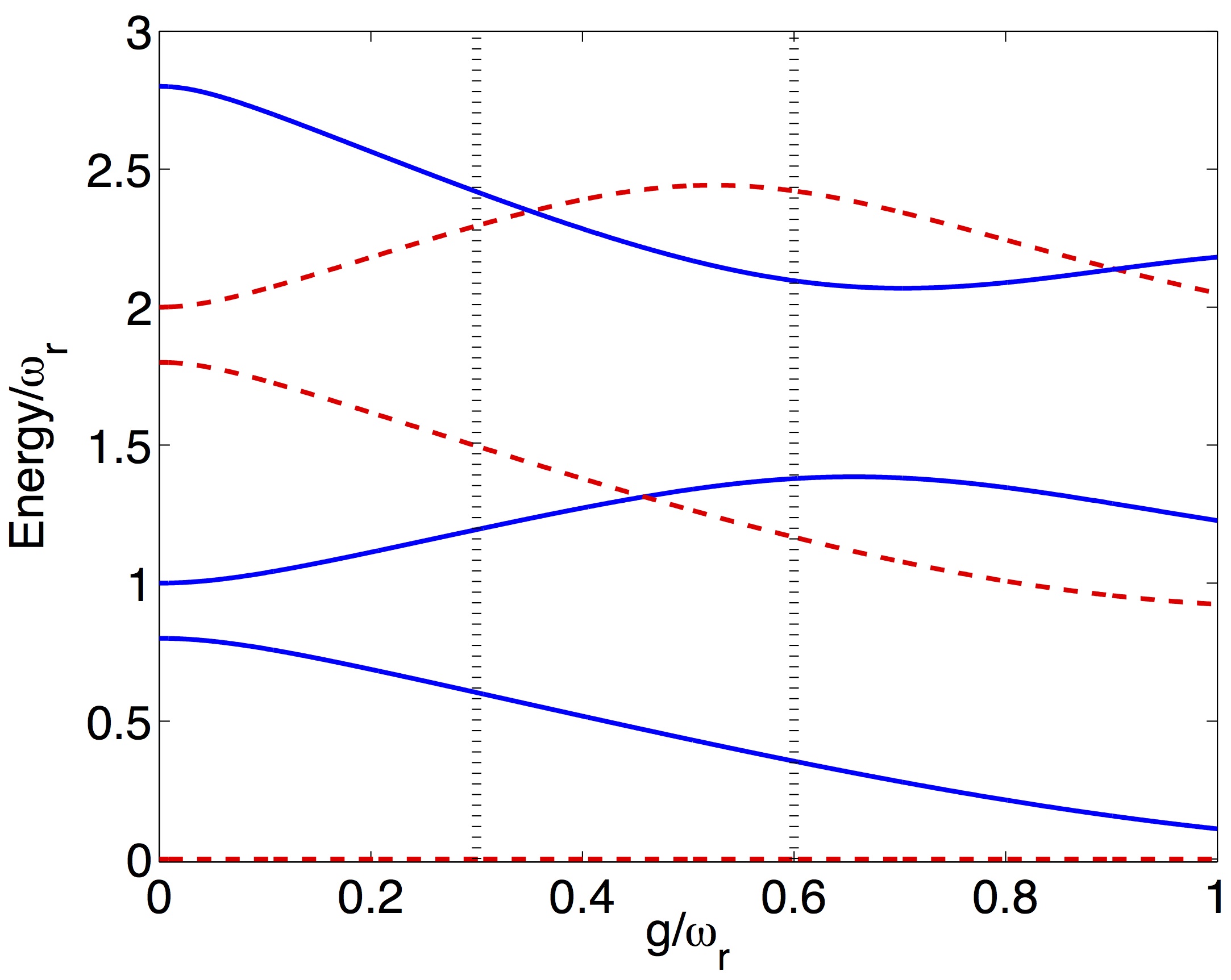}
\caption{{\bf Spectrum of the quantum Rabi model.} Energy levels of the quantum Rabi model as a function of the dimensionless parameter $g/\omega_r$. We assume $\hbar=1$. Parameter values are expressed in units of $\omega_r$ and we consider a detuned system qubit $\omega/\omega_r=0.8$. Energies are rescaled in order to set the ground level to zero. The parity of the corresponding eigenstates is identified, blue continuous line for odd and red dashed lines for even states.}
\label{rabispectrum}
\end{figure}

The quantum Rabi model (QRM)~\cite{Rabi36,Braak2011} describes the dipolar coupling of a two-level system and a single-mode cavity field, as described by the Hamiltonian
\begin{equation}
\label{Hrabi}
H_{S} = \hbar\omega_r a^\dagger a + \frac{\hbar\omega}{2} \sigma_z + \hbar g \sigma_x \left( a^\dagger + a \right),
\end{equation}
where $a^\dagger$($a$) represents the creation(annihilation) operator of the cavity field, while $\sigma_x$ and $\sigma_z$ are Pauli operators defined in the qubit Hilbert space. We denote, in Eq.~(\ref{Hrabi}), the cavity mode frequency, $\omega_r$, the qubit frequency spacing, $\omega$, and the interaction strength, $g$. If we restrict ourselves to near resonant interactions, $\omega \approx \omega_r$, depending on the parameter $g/\omega_r$, two regimes can be identified: the SC regime for $g/\omega_r\ll1$ and the USC regime
for  $0.1\lesssim g/\omega_r\lesssim1$. In the former, the Hamiltonian of Eq.~\eqref{Hrabi} reduces to the celebrated Jaynes-Cummings model~\cite{JCmodel}, where the conservation of the excitation number $\hat{N}=a^\dagger a + \sigma_z $ turns the model analytically solvable. On the contrary, in the USC regime, the field and the qubit merge into polariton states that feature a discrete symmetry $Z_2$, see Fig.~\ref{rabispectrum}. This symmetry is characterized by the parity operator $\hat{\Pi}_S = -\sigma_z\ e^{i\pi a^\dagger a }$, such that $\hat{\Pi}_S |\psi_{j}\rangle=\pm|\psi_{j}\rangle$ with $j=0,\hdots, \infty$. Here, we denote $\ket{\psi_{j}}$ as polariton eigenstates of energy $\hbar\omega_{j}$. Furthermore, this parity symmetry turns the model solvable~\cite{Braak2011}, and approximations exist in limiting cases, as is the case of the perturbative USC regime~\cite{Irish2007} and the deep strong coupling (DSC) regime~\cite{Casanova2010,DeLiberato2014}.

We consider the QRM in the USC regime plus an ancillary qubit interacting with the cavity field,
\begin{equation}
\label{Hfull}
H = H_{S} + H_{A}\ ,\quad  H_{A} = \frac{\hbar\omega_a}{2} \sigma^a_z + \hbar g_a   \sigma^a_x \left( a^\dagger + a \right).
\end{equation}
Later, we will assume that the frequency $\omega_a$ can be tuned in real time, a requirement that can be fulfilled in superconducting circuits~\cite{Koch2007, Srinivasan2011}. We set the ancilla-cavity field  interaction $g_a$ to be in the SC regime. However, counterintuitively, we will show that the presence of the USC system activates the counter-rotating terms of the ancilla $ g_a \left( \sigma^a_- a + \sigma^a_+ a^\dagger \right)$ even for small $g_a/\omega_r$. Indeed, the relevance of the ancilla counter-rotating terms depends on the polariton eigenstate more than on the ratio $g_a/\omega_r$, as long as the interaction between the ancillary qubit and the USC system is in the SC regime. Here, $\sigma^a_\pm =\left( \sigma^a_x \pm i\ \sigma^a_y)\right/2$ is the raising(lowering) operator of the ancilla.

\begin{figure}[]
\centering
\includegraphics[width=0.98\columnwidth]{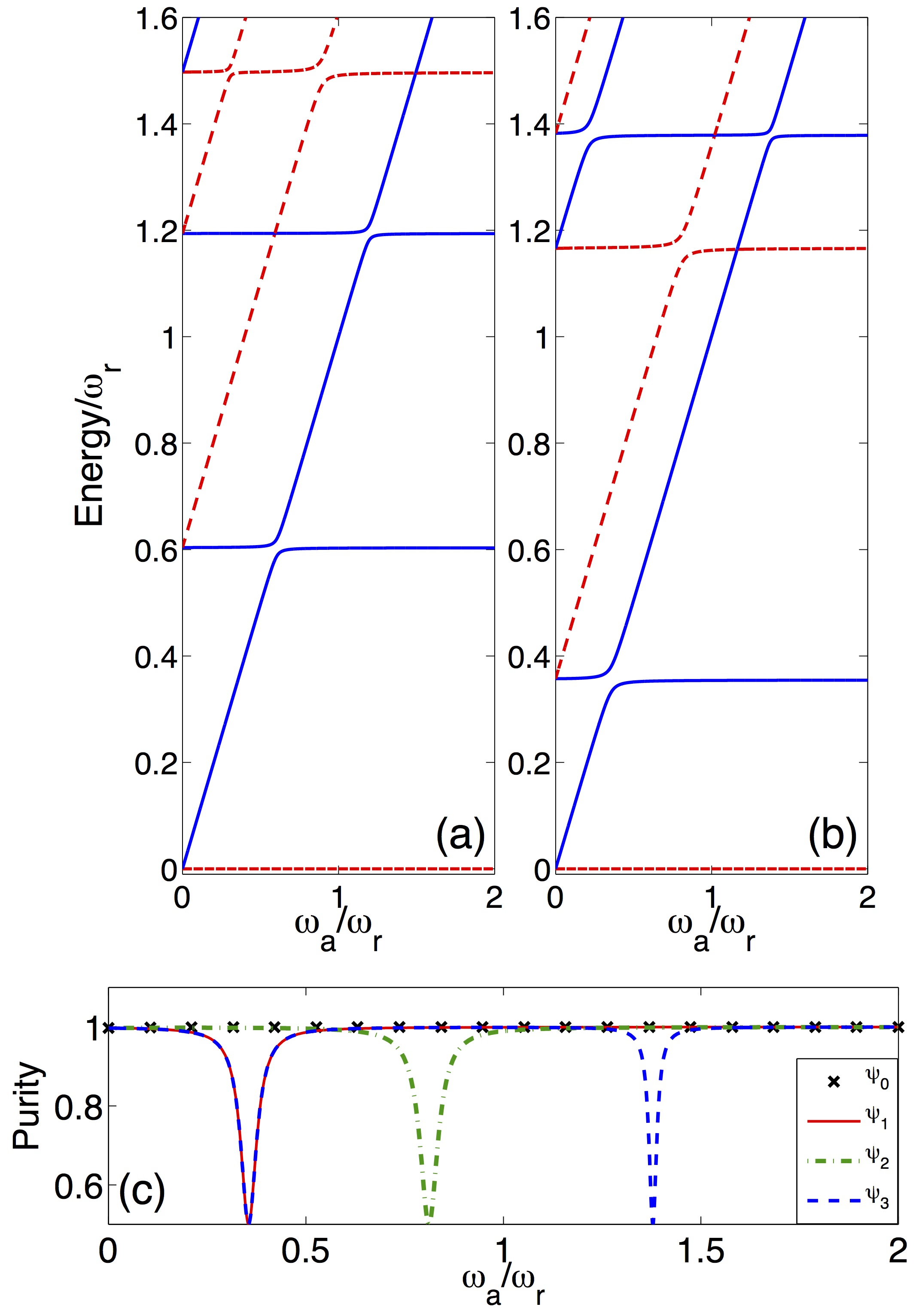}
\caption{{\bf Global spectral properties as a function of the ancilla energy spacing.} (a), (b) Energy levels of the full model of Eq.~\eqref{Hfull} as a function of the ancilla frequency $\omega_a$. We assume 
$\hbar=1$.  For both (a) and (b), the USC qubit frequency is $\omega/\omega_r = 0.8$ and the ancilla-field cavity interaction strength is $g_a/\omega_r=0.02$. The USC qubit coupling is $g/\omega_r =0.3$ for (a) and $g/\omega_r =0.6$ for (b). Energies are rescaled in order to set the ground level to zero. The global parity of the corresponding eigenstates is identified, blue continuous line for odd and red dashed lines for even states.
(c) Purity $\mathcal{P}$ of the reduced density matrix of the ancillary qubit for different global system eigenstate, as a function of the ancilla frequency. For the ground state $\ket{\psi_0}$, $\mathcal{P} $ is always unity.}
\label{fullspectrum}
\end{figure}

The spectrum of the full Hamiltonian \eqref{Hfull} is shown in Fig.~\ref{fullspectrum}(a) (Fig.~\ref{fullspectrum}(b)) as a function of the ancilla frequency for $g/\omega_r=0.3$ ($g/\omega_r\!=\!0.6$), corresponding to the vertical lines displayed in Fig.~\ref{rabispectrum}. The total ancilla-system spectrum, associated to Hamiltonian $H$ in Eq.~(\ref{Hfull}), presents three main features. Firstly, the system still preserves the $Z_2$ symmetry with the global parity operator $\hat{\Pi} = \sigma^a_z\otimes\sigma_z\ e^{i\pi a^\dagger a }=\hat{\Pi}_A\otimes\hat{\Pi}_S$. Notice that eigenstates with global parity $+1(-1)$ are represented by dashed-red(continuous-blue) lines in Figs.~\ref{fullspectrum}(a) and~\ref{fullspectrum}(b). Secondly, introducing the ancillary qubit results in the splitting of the energy levels of polaritons. There are regions where the energy differences behave linearly with $\omega_a/\omega_r$, so the main contribution of the ancilla comes from its self-energy. This behavior can be explained if we consider the average value of the  quadrature $\hat{X} = a +a^\dagger$ appearing in the cavity-ancilla interaction of Eq.~(\ref{Hfull}). It vanishes for diagonal projections in the polariton basis, that is, $\langle\psi_{j}|\hat{X}|\psi_{j}\rangle=0$ for $j=0,\hdots, \infty$ (see Supplementary information). Thirdly, intersections between levels of different global parity subspaces show that those eigenstates are not coupled. On the contrary, avoided crossings between eigenenergies sharing the same global parity confirm that such states experience a direct coupling. In the following, we will show how this feature allows for selective state engineering of the USC polaritons.

In order to use the ancillary qubit as a tool to characterize and to measure polaritons in the USC regime, the ground state of the ancilla plus USC system must be separable. This condition is fulfilled as seen in Fig.~\ref{fullspectrum}(c), where we show the purity $\mathcal{P}  =  \mathrm{Tr}\left\{ \rho_a^2 \right\} $ for the ground and first excited states. We define the ancilla reduced density matrix as $\rho_a = \mathrm{Tr}_{\rm{polariton}} \left\{ \rho \right\} $, where the partial trace runs over the USC system degrees of freedom. If $\mathcal{P}=1$, the ancilla and the polariton are in a separable state. Contrariwise, in coincidence with avoided crossings in the spectrum, see Fig.~\ref{fullspectrum}(a,b), the purity presents some dips for excited states revealing ancilla-system entanglement, with $\mathcal{P}=1/2$ corresponding to a maximally entangled state.

\section{Real-time dynamics and spectroscopic protocol}
 Let us now analyze the total system real-time dynamics. From the previous consideration on the spectrum, it emerges that we can describe the ancilla and the USC system in the Hilbert space denoted by the tensor product of both subsystems $\mathcal{H}=\mathcal{H}_{\rm ancilla}\otimes\mathcal{H}_{\rm polariton}$. Accordingly, we formally rewrite the Hamiltonian of Eq.~\eqref{Hfull} as
\begin{eqnarray}
H = \hbar\sum_{j} \omega_{j} \ket{\psi_{j}}\bra{\psi_{j}} + \frac{\hbar\omega_a}{2}\sigma_z^a + H_I \nonumber\\
H_I = \hbar g_a \sigma_x^a\sum_{ij}\left[ k_{ij}  \ket{\psi_i}\bra{\psi_j} + k^*_{ji}  \ket{\psi_i}\bra{\psi_j}  \right],
\end{eqnarray}
where we denote $\ket{\psi_j}$ as polariton states of energy $\hbar\omega_j$. These states are given by eigenstates of $H_S$~\eqref{Hrabi} and we also define the transition matrix elements $k_{ij}=  \bra{\psi_i}a\ket{\psi_j}$.

Because of the strong anharmonicity of the QRM, when the ancilla frequency matches a given polariton transition $\omega_a= \omega_\alpha-\omega_\beta$, we can perform a RWA and rewrite the interaction Hamiltonian $H_I$ as (see Supplementary information)
\begin{equation}
H_I = \hbar g_a \left( k_{\alpha\beta} + k^*_{\beta\alpha} \right) \sigma_-^a\  \ket{\psi_\alpha}\bra{\psi_\beta}  + \textrm{H.c.},
\label{Hint}
\end{equation}
where we fixed $\omega_\alpha> \omega_\beta $.
Such a Hamiltonian induces coherent excitation transfer between the ancilla qubit and the polariton system. Notice that the matrix element $k_{ij}$ is non-vanishing only for transitions that link states of opposite parity in the polaritonic system. To check the validity of our analytical treatment, we simulate the real-time dynamics of the full model. We take into account decoherence effects by means of second-order time-convolutionless projection operator method~\cite{OpenQBook}, which correctly describes the dissipative dynamics in the USC regime. In this simulation we have considered zero-temperature thermal baths and noises acting on the $\hat{X}$ quadrature and transversal noise ($\sigma_x$) for both two-level systems. Realistic parameters for superconducting circuits have been considered. Fig.~\ref{spectroscopy}(c) shows an example of Rabi oscillations (green continuous line) between the states $\ket{e}\ket{\psi_0}$ and $\ket{g}\ket{\psi_1}$, where we denoted with $\ket{g}$($\ket{e}$) the ground(excited) state of the ancillary qubit. 

We stress that counter-rotating terms $ g_a \left( \sigma^a_+ a^\dagger + \sigma^a_- a \right)$ of the ancilla-cavity coupling, see Eq.\eqref{Hfull}, play an important role in the total system dynamics. Those terms contribute to Eq.~{\eqref{Hint} with the coefficients  $k_{\alpha\beta}$ and $k^*_{\beta\alpha}$, given that we fixed $\omega_\alpha> \omega_\beta $. Their effect is highlighted in Fig.~\ref{spectroscopy}(c) by reproducing the same dynamics in the case in which  such contributions are artificially neglected (black dashed line). Notice that, if the system qubit were removed, or if it were interacting in the SC regime, the effect of counter-rotating terms of the ancilla-cavity interaction would be negligible for such small values of the ratio $g_a/\omega$. In fact, the presence of a qubit in the USC regime modifies the mode structure of the cavity field in such a way that the coefficients $k_{ij}$ can be non-vanishing also for $\omega_i>\omega_j$ (see Supplementary information). In simple words, this condition implies that removing a photon results in an increase of the system energy, in striking contrast with Jaynes-Cummings-like energy spectrum. 
Indeed, the counterintuitive breaking of the RWA, explained here for an ancilla interacting with a ultrastrongly coupled system, unveils a general feature of the USC regime.

The expectation value of $\sigma^a_z$ can be measured by detuning $\omega_a$ out of resonance, with respect to the USC system, and in resonance with an idle cavity for readout~\cite{Gambetta2007, Lad2007}.
This enables us to design a spectroscopy protocol  for the USC system, which identifies the parity of each energy level.
Such a protocol consists in keeping track of the expectation value $\langle \sigma^a_z \rangle$ during the time-evolution, after initializing the USC system in its ground level and the ancilla in its excited state $\ket{\phi_0^\textrm{e}} =  \ket{\psi_0}\ket{e}$. Notice that the ground and first-excited states of the QRM Hamiltonian have even and odd parity, respectively.
The initialization can be realized when the ancilla is far off-resonance, then its frequency can be suddenly switched~\cite{Srinivasan2011} to be within the relevant frequency range. 
As the ancilla frequency becomes closer to a given transition of the USC system, the amplitude of the excitation transfer increases, granted that the process preserves the global parity.
Thus, sampling the ancilla dynamics for different values of $\omega_a$, we can deduce the USC system eigenvalues belonging to a specific parity subspace (blue continuous line in Fig.~\ref{spectroscopy}a and Fig.~\ref{spectroscopy}b). We define the visibility as half the difference between the maximum and the minimum values reached by $\langle \sigma^a_z \rangle$ during its time-evolution.
Considering realistic parameters of superconducting circuit technology, taking $\omega_r=2\pi\times5$~GHz, the first three resonance peaks can be obtained within a time of approximatively 10~$\mu s$ (see Supplementary information).

\begin{figure}[]
\centering
\includegraphics[width=0.98\columnwidth]{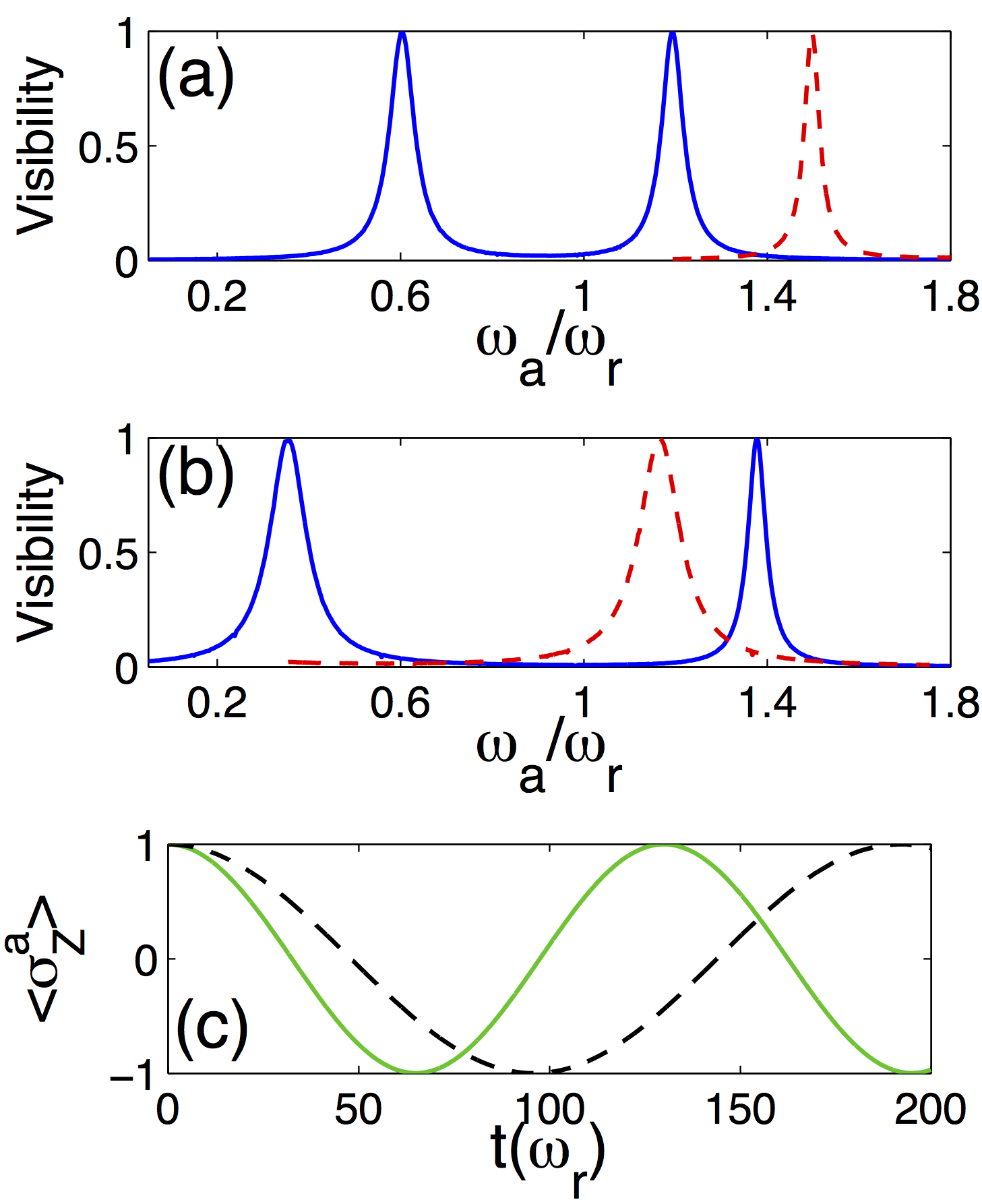}
\caption{{\bf Spectroscopic protocol and real time dynamics.} (a), (b) Numerical simulation of the spectroscopy protocol. Visibility of the ancilla population oscillations as a function of frequency $\omega_a$. Physical parameters correspond to the vertical cuts in Fig.~\ref{rabispectrum}. For both (a) and (b), the system qubit frequency is $\omega/\omega_r= 0.8$ and the ancilla-field cavity coupling is $g_a/\omega_r=0.02$. The USC system coupling is $g/\omega_r =0.3$ for (a) and $g/\omega_r =0.6$ for (b). The parity of each energy level is identified,  blue continuous line for odd and red-dashed lines for even states. (c) Comparison of full model (green continuous line) to the dynamics obtained when removing counter-rotating terms from the ancilla-cavity interaction (black dashed line). System parameters are the same that in box (b). In all cases, decay rates are $\gamma/\omega_r = 10^{-3}$ for the system qubit,  $\gamma_r/\omega_r =10^{-4}$ for the cavity and $\gamma_a/\omega_r = 10^{-4}$ for the ancilla.}
\label{spectroscopy}
\end{figure}

 In the same way, we can obtain the level structure of the \textit{even} subspace (red dashed  line in Fig.~\ref{spectroscopy}a and Fig.~\ref{spectroscopy}b) by repeating the protocol with the odd initial state $\ket{\phi_0^\textrm{o}} = \ket{\psi_1}\ket{e}$, i.e., both the ancilla and the USC system in their first-excited state. The total system can be initialized in such a state via state-transfer process (see below) plus a spin-flip operation on the ancilla qubit. The proposed spectroscopic protocol allows us to obtain the parity structure of the USC system in a direct way. Hence, one could check the eigenstate-parity inversion (see Fig.~\ref{rabispectrum}), which is specific to the QRM and represents a distinctive signature of the USC regime. Higher energy levels can be obtained in a similar way with a multi-step procedure. Notice that the widths of the resonance peaks in Fig.~\ref{spectroscopy} are proportional to the matrix elements $k_{ij}$, hence they contain information about the eigenstates of the USC system.

\section{Tomography and state engineering}  So far we have considered the ancillary qubit dynamics as a tool to investigate the spectral structure of the USC system. Let us now focus on how this ancilla can be used as a tool to fully measure and control the USC, granted that a limited number of its eigenstates can be excited. First, we show how the tomography of the ancillary qubit~\cite{Steffen2006} enables us to recover all the coefficients of the USC density matrix. The protocol to be followed consists in initializing the ancilla in a proper state, implementing a selective state transfer between the USC system and the ancilla, and performing tomography of the latter. After the initialization of the ancilla, the global density matrix reads $\rho = \sum_{i,j}\rho_{ij}\ket{\psi_i}\bra{\psi_j}\otimes \ket{g}\bra{g}$.
For opposite parity eigenstates  $\ket{\psi_n}$ and $\ket{\psi_m}$, implementing the state transfer process $ \ket{\psi_n}\ket{g} \leftrightarrow \ket{\psi_{m}}\ket{e}$, and tracing over the USC system degrees of freedom, we obtain the ancillary qubit density matrix
\begin{equation}
\rho_a = \rho_0\ket{g}\bra{g} + \rho_{nn} \ket{e}\bra{e} + \rho_{nm}\ket{e}\bra{g} + \rho_{mn}\ket{g}\bra{e},
\end{equation}
where $\rho_0 = \sum_{i\neq n}\rho_{ii}$. Hence, performing tomography over the ancilla yields the value of the population in state $\ket{\psi_n}$ and the coherence coefficients with $\ket{\psi_m}$.
In order to infer the coherences between USC system states of identical parity, a slightly different procedure must be used. In this case, a two-step state transfer process can be implemented, making use of a third level of opposite parity to mediate the interaction (see Supplementary information). Then, iterating the protocol for all couples of relevant eigenstates, the complete density matrix of the USC system state can be reconstructed. 

Notice that the selective state-transfer processes introduced for the tomography protocol can be performed in a reverse way to engineer the state of the USC system itself. Assuming that any single-qubit gate can be performed on the ancilla, the components of the USC system state in the energy eigenbasis can be individually addressed by means of the selective interactions of Eq.~\eqref{Hint}. Parity-forbidden transitions can be circumvented by means of a two-step protocol (see Supplementary information). For instance, the USC system can be prepared in any superposition of its eigenstates by iteratively initializing the ancilla qubit in the desired state, tuning its energy spacing to match a given transition, and performing a selective state-transfer.
This feature can be exploited in order to connect ultrastrongly-coupled systems with standard quantum information processing devices. For instance a logical qubit can be encoded in the ancilla state and then transferred to the polariton, where the computational benefits of the USC coupling can be exploited~\cite{Nataf2011,Romero2012,Kyaw2014}.

\section{Discussion}
In conclusion, we have analyzed the interaction between an ancillary qubit and an ultrastrongly coupled qubit-cavity system. We find that the presence of a USC qubit- cavity system modifies the interaction of the cavity with the ancillary qubit in a nontrivial manner. We have designed a spectroscopy protocol able to detect parity-inversion of eigenstates, a signature of the USC regime in the QRM, requiring control over a single ancillary qubit and tunability of its effective frequency. The present method can be applied in order to certify that a device is operating in the USC regime of the quantum Rabi model. Moreover, we show that the same ancilla may be used as a tool to engineer the dynamics of arbitrary USC system states. 
The proposed method overcomes the lack of decoupling mechanisms in the USC regime, requiring minimal external resources. Our results pave the way to novel applications of the USC regime of the QRM in quantum technologies and quantum information processing.

\vspace{0.2 cm}
{\setlength{\parindent}{0pt}
\begin{acknowledgments}
This work was supported by the Spanish MINECO FIS2012-36673-C03-02; Basque Government IT472- 10; UPV/EHU UFI 11/55 and FONDECYT 1150653 CCQED, PROMISCE, and SCALEQIT European projects. \end{acknowledgments}
}

\begin{widetext}

\section*{SUPPLEMENTAL MATERIAL}

\section{Some properties of the Rabi model}
\label{sec1}
We will not provide here a detailed description of the properties of the Rabi model eigenstates (instead we refer to~\cite{Braak2011_sup}). The main feature we will focus on is the parity conservation of the Rabi Hamiltonian. Let us define the parity operator $\Pi=-\sigma_z e^{i\pi a^\dagger a}$, which corresponds to the parity of the number of excitations in the composite system. Once the parity operator is defined, it is straightforward to show that it commutes with the Rabi Hamiltonian. This ensures that any eigenstate of the Rabi model is also an eigenstate of the parity operator.

Let us now consider the action of the creation operator on an arbitrary eigenstate $\ket{\psi_n}$. From its very definition, the creation operator creates one excitation inside the cavity, thus bringing $\ket{\psi_n}$ to a vector of opposite parity. In other words, $a^\dagger\ket{\psi_n}$ belongs to a subspace orthogonal to the one in which lies $\ket{\psi_n}$. Eventually we have shown the following relation:
\begin{equation}
\bra{\psi_j}a^\dagger\ket{\psi_n}=0
\end{equation}
for any $j$ such that the parity of $\ket{\psi_j}$ is the same as $\ket{\psi_n}$, as shown in Fig.\ref{transitionmatrix}.  This demonstration naturally extends to the annihilation operator $a$. As a corollary, we have:
\begin{equation}\label{Corro}
\bra{\psi_n}a^{(\dagger)}\ket{\psi_n}=0
\end{equation}

We show in the main text that apart from avoided crossings, the full model eigenstates are in product states made of eigenstates of the Rabi model and the ancilla being in the ground or excited state. Eq.~\eqref{Corro} proves that the only contribution from the ancilla to the eigenenergies comes from the free Hamiltonian $(\omega_a/2)\sigma_z^a$. This explains the behavior of the eigenenergies as a function of $\omega_a$: fully degenerate at $\omega_a=0$, then increasing linearly.

\begin{figure}[h]
\centering
\includegraphics[width=0.3\textwidth]{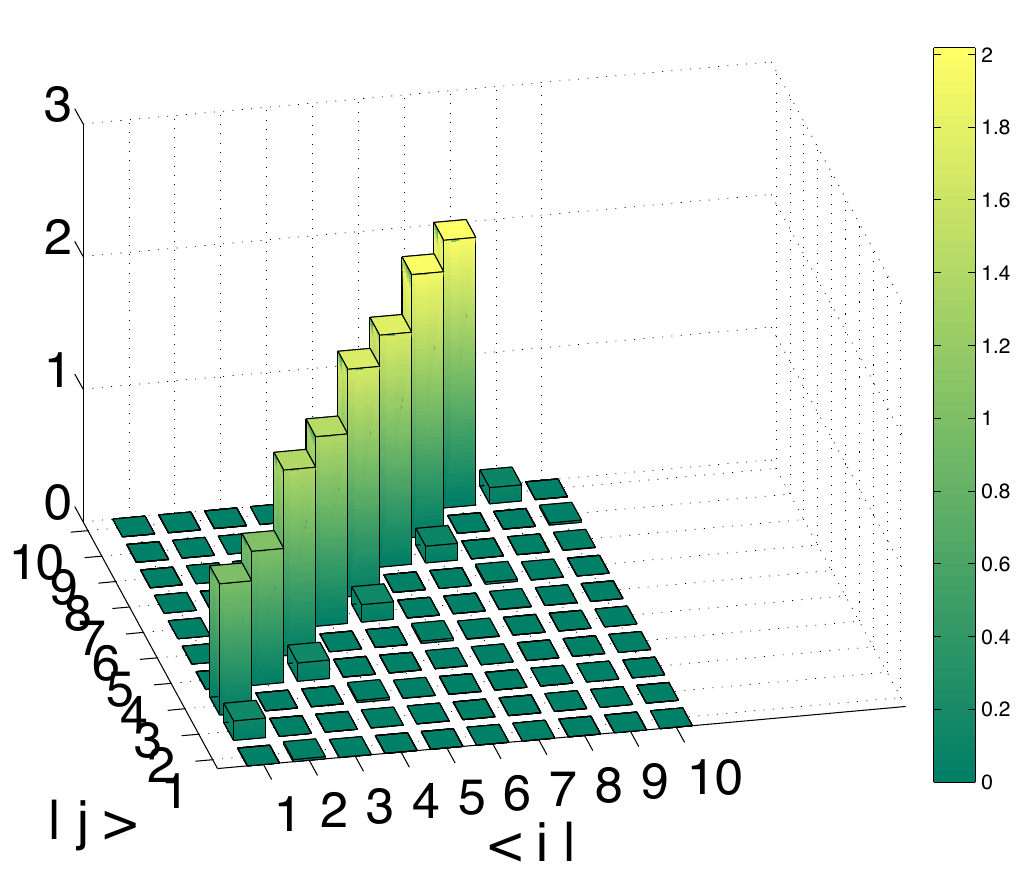}\hspace{2cm}
\includegraphics[width=0.3\textwidth]{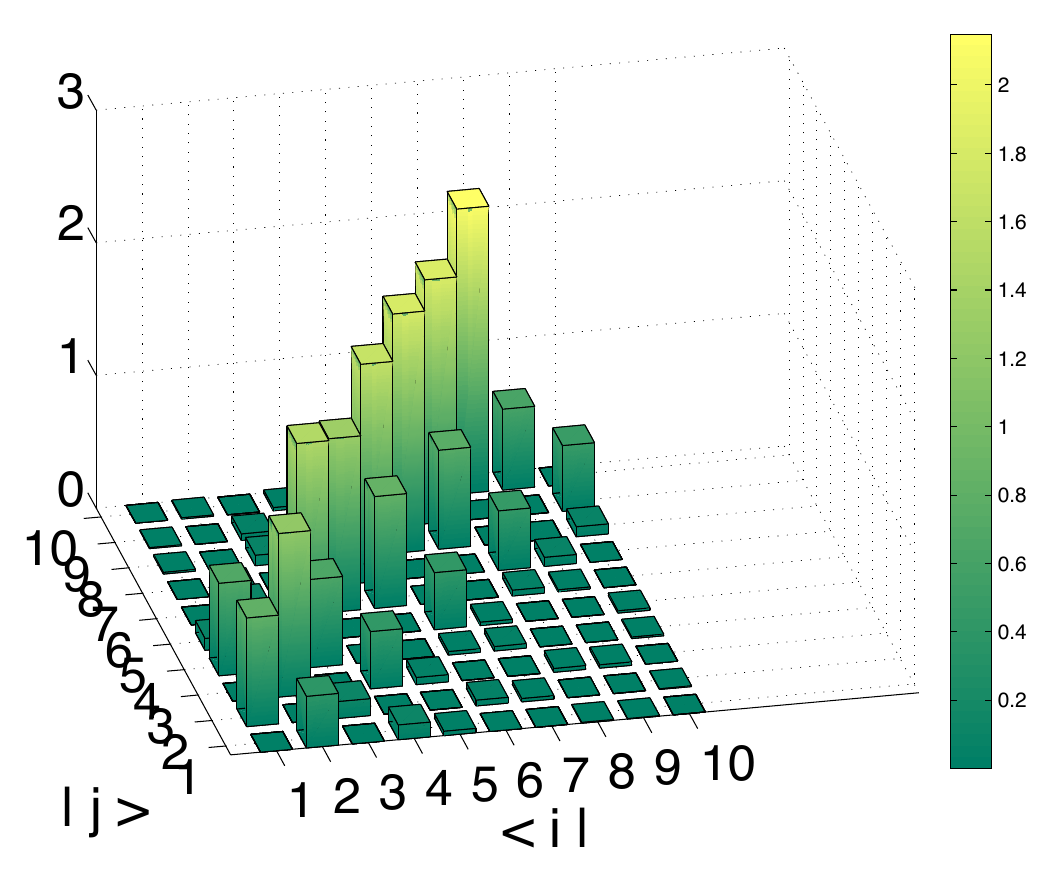}
\caption{Absolute values of the elements of the transition matrix $k_{ij} = \bra{\psi_i} a \ket{\psi_j}$. The left box corresponds to the strong coupling regime ($g/\omega_r = 0.03$), while the right box corresponds to the USC regime ($g/\omega_r = 0.6$), i.e. the Rabi model. The diagonal term vanish, as shown in section \ref{sec1}. Notice that in the SC regime, where the Jaynes-Cummings model applies, the coefficients $k_{ij}$ vanish for $\omega_i > \omega_j$.}
\label{transitionmatrix}
\end{figure}

\section{Derivation of the effective interaction Hamiltonian}
We consider the system as being composed of the the Rabi model interacting via the resonator field with the ancillary qubit. This means the total Hamiltonian reads:
\begin{eqnarray}
\mathcal{H} = \sum_n \omega_n \ket{\psi_n}\bra{\psi_n} + \frac{\omega_a}{2}\sigma_z^a + \mathcal{H}_I \nonumber\\
\mathcal{H}_I = g_a\sigma_x^a(a+a^\dagger)
\end{eqnarray}
where we denoted with $\ket{\psi_n}$ the eigenstates, of increasing energy $\omega_n$, of the Rabi model. Using the completeness relation $\mathcal{I}=\sum_n\ket{\psi_n}\bra{\psi_n}$, the interaction Hamiltonian becomes:
\begin{equation}
\label{Hbase}
\mathcal H_I=g_a\sigma_x^a\sum_{i,j}k_{ij}\ket{\psi_i}\bra{\psi_j}+k_{ji}^*\ket{\psi_i}\bra{\psi_j}
\end{equation}
where $k_{ij}=\bra{\psi_i}a\ket{\psi_j}$. In the previous section we recalled that for any $i$ we have $\bra{\psi_i}a+a^\dagger\ket{\psi_i}=0$. Thus we can order the double sum in equation Eq.~\eqref{Hbase} to get:
\begin{equation}
\mathcal H_I=g_a\sigma_x^a\sum_{i>j}(k_{ij}+k_{ji}^*)\ket{\psi_i}\bra{\psi_j}+(k_{ji}+k_{ij}^*)\ket{\psi_j}\bra{\psi_i}
\end{equation}
Since the $\ket{\psi_n}$'s are labeled in increasing energy, we can interpret the two operators in the sum the following way: one raising the energy of the polariton, $\ket{\psi_i}\bra{\psi_j}$, the other one lowering the energy $\ket{\psi_j}\bra{\psi_i}$. Now we will assume that the spectrum of the Rabi model is non-linear enough so that we are able to isolate one particular transition frequency $\omega_{ij}=\omega_i-\omega_j>0$. This anharmonic assumption is valid in the regime of $g/\omega_r\lesssim2$, which is the one we consider here. Thus we can perform a new Rotating Wave Approximation (RWA) when bringing the frequency of the ancilla close to resonance with $\omega_{ij}$. More precisely, we move $\mathcal H_I$ to the interaction picture:
\begin{equation}
\tilde{\mathcal H}_I(t)=g_a(\sigma_+^ae^{i\omega_at}+\sigma_-^ae^{-i\omega_at})((k_{ij}+k_{ji}^*)\ket{\psi_i}\bra{\psi_j}e^{i\omega_{ij}t}+(k_{ji}+k_{ij}^*)\ket{\psi_j}\bra{\psi_i}e^{-i\omega_{ij}t})
\end{equation}
In this expression we identify two oscillating frequencies: $\omega_a+\omega_{ij}$ and $\omega_a-\omega_{ij}\equiv\delta$. In this context we will perform the standard RWA, neglecting the quickly oscillating terms. The interaction picture Hamiltonian reads:
\begin{equation}
\tilde{\mathcal H}_I(t)=g_a((k_{ij}+k_{ji}^*)\sigma_-^a\ket{\psi_i}\bra{\psi_j}e^{-i\delta t}+(k_{ji}+k_{ij}^*)\sigma_+^a\ket{\psi_j}\bra{\psi_i}e^{i\delta t})
\end{equation}
thus yielding a Jaynes-Cummings-like interaction Hamiltonian.

\section{Estimation of the time required to perform the spectroscopy protocol}

Our protocol allows for analyzing the spectrum of the polariton, based on measurements performed on the ancillary qubit. This means the relevant parameters for this protocol are well-known and the manipulations are now standard~\cite{Srinivasan2011_sup}. In this section we will provide a rough estimation of the time required to detect the peaks in Fig. 4 of the main text.

The right order of magnitude for the experimental $\omega_2$ spacing is given by the full width at half maximum (FWHM) of those peaks. In our case, the FWHM is of the order of $0.1\omega_r$. Besides, we span with $\omega_r$ an interval of approximate length $2\omega_r$. Considering 5 points per peak, we obtain an upper bound on the number of points we want to measure of 100. This value could be further reduced performing a more clever analysis of the spectrum.

Every point actually corresponds to computing the visibility of Rabi oscillations at a given ancilla frequency. This can be done by measuring the ancilla until half a period, in other words by monitoring the ancilla for a time $T_{\mathrm{half}}\approx50/\omega_r$ (see Fig. 4c in the main text). This monitoring requires to measure $\sigma_z^a$ roughly 50 times, every measurement being of a duration at most $T_{\mathrm{half}}$. For a standard cavity in circuit QED, we have $\omega_r\approx2\pi\times5$ GHz, which gives approximately 100 ns to recover the visibility at a given frequency $\omega_2$.

Going from one point to another means tuning the ancillary qubit frequency. This can be done in a few nanoseconds ~\cite{Srinivasan2011_sup}, hence it is negligible compared to the computation of a single point. In the end, summing 100 ns for 100 values of $\omega_2$, the whole spectroscopy duration is of the order of 10 microseconds.

\section{Multi-step process for tomography}

The goal of this section is to show how one may address transitions that should be forbidden because of parity conservation. Let $\ket{\psi_a}$ and $\ket{\psi_b}$ two eigenstates of the Rabi model of same parity. We want to distinguish the pure states made of an arbitrary superposition of $\ket{\psi_a}$ and $\ket{\psi_b}$ from the statistical mixture with same weights. To this end, we will make use of an auxiliary eigenstate of opposite parity $\ket\phi$. In the following we will consider two different cases: first when the energy of $\ket\phi$ lies in between the energies of $\ket{\psi_a}$ and $\ket{\psi_b}$, then when it doesn't.

\subsection{Forbidden transition with an auxiliary one in between}
We suppose here the following ordering of the Rabi eigenstates energies. Namely: $E_{\ket{\psi_a}}<E_{\ket{\phi}}<E_{\ket{\psi_b}}$. We will consider that only those three levels are populated, assuming that we are able to initialize the ancilla in the $\ket+\equiv(\ket e+\ket g)/\sqrt2$ state. The initial state reads:
\begin{equation}
\ket{\phi_i}=(\alpha\ket{\psi_a}+\beta\ket{\psi_b}+\gamma\ket\phi)\ket +
\end{equation}
First we perform half a Rabi oscillation between $\ket{\psi_a}$ and $\ket\phi$. This transforms the global state as follows:
\begin{equation}
\ket{\phi_1}=\alpha\frac{\ket{\psi_a}\ket g+\ket\phi\ket g}{\sqrt2}+\beta\ket{\psi_b}\ket++\gamma\frac{\ket{\psi_a}\ket e+\ket\phi\ket e}{\sqrt2}
\end{equation}
Then we repeat the protocol for the transition between $\ket\phi$ and $\ket{\psi_b}$. Thus we have:
\begin{equation}
\ket{\phi_2}=\alpha\frac{\ket{\psi_a}\ket g+\ket\phi\ket g}{\sqrt2}+\beta\frac{\ket{\phi}\ket e+\ket{\psi_b}\ket e}{\sqrt2}+\gamma\frac{\ket{\psi_a}\ket e+\ket{\psi_b}\ket g}{\sqrt2}
\end{equation}
which we can write in a more convenient way:
\begin{equation}
\ket{\phi_2}=\frac1{\sqrt2}(\ket{\phi_a}\ket g+\ket{\phi_b}\ket e)
\end{equation}
where $\ket{\phi_a}$ and $\ket{\phi_b}$ are two non-orthogonal states. Eventually, the reduced density matrix of the ancillary qubit reads:
\begin{equation}
\rho_a=\left(\begin{matrix}\vert\alpha\vert^2+\frac12\vert\gamma\vert^2 & \frac12(\alpha\beta^*+\alpha\gamma^*+\gamma\beta^*)\\ \frac12(\alpha^*\beta+\alpha^*\gamma+\gamma^*\beta) & \vert\beta\vert^2+\frac12\vert\gamma\vert^2\end{matrix}\right)
\end{equation}
Finally, assuming that we can infer the coherences between $\ket{\psi_a}$ and $\ket\phi$ and between $\ket{\psi_b}$ and $\ket\phi$ separately, this protocol allows for measuring the coherence terms relative to a forbidden transition -- which correspond here to the product $\alpha\beta^*$.

\subsection{Forbidden transition between two consecutive eigenstates}
In the case where the forbidden transition involves two consecutive eigenstates the result is a bit different. The energies correspond to $E_{\ket{\psi_a}}<E_{\ket{\psi_b}}<E_{\ket{\phi}}$. The initial state is again
\begin{equation}
\ket{\phi_i}=(\alpha\ket{\psi_a}+\beta\ket{\psi_b}+\gamma\ket\phi)\ket +
\end{equation}
The first excitation transfer, associated with the transition between $\ket{\psi_a}$ and $\ket\phi$ yields:
\begin{equation}
\ket{\varphi_1}=\alpha\frac{\ket{\psi_a}\ket g+\ket\phi\ket g}{\sqrt2}+\beta\ket{\psi_b}\ket++\gamma\frac{\ket{\psi_a}\ket e+\ket\phi\ket e}{\sqrt2}
\end{equation}
Then when it comes to the transition between $\ket{\psi_b}$ and $\ket\phi$ the global state becomes:
\begin{equation}
\ket{\varphi_2}=\alpha\frac{\ket{\psi_a}\ket g+\ket{\psi_b}\ket e}{\sqrt2}+\beta\frac{\ket{\phi}\ket e+\ket{\psi_b}\ket e}{\sqrt2}+\gamma\frac{\ket{\psi_a}\ket e+\ket\phi\ket e}{\sqrt2}
\end{equation}
which we can once again write, defining two non-orthogonal states $\ket{\varphi_a}$ and $\ket{\varphi_a}$ different than before:
\begin{equation}
\ket{\varphi_2}=\frac1{\sqrt2}(\ket{\varphi_a}\ket g+\ket{\varphi_b}\ket e)
\end{equation}
and for the reduced density matrix of the ancilla:
\begin{equation}
\rho_a=\left(\begin{matrix}\vert\beta\vert^2+\frac12\vert\alpha\vert^2 & \frac12(\alpha\gamma^*+\beta\alpha^*+\beta\gamma^*)\\ \frac12(\alpha^*\gamma+\beta^*\alpha+\beta^*\gamma) & \vert\gamma\vert^2+\frac12\vert\alpha\vert^2\end{matrix}\right)
\end{equation}
Once again, one may access the coherence terms relative to the forbidden transition.

\end{widetext}


\begin{thebibliography}{99}

\bibitem{Walther2006} 
Walther, H.,Varcoe,  B. T. H., Englert, B.-G. \&  Becker, T. Cavity quantum electrodynamics.
\textit{Rep. Prog. Phys.} \textbf{69}, 1325 (2006).

\bibitem{QuantumBook} 
Haroche, S. \&  Raimond, J.-M. Exploring the Quantum.
\textit{Oxford University Press}, New York, 2006. 

\bibitem{Rabi36}
Rabi, I. I. On the Process of Space Quantization.
\textit{Phys. Rev.} {\bf 49}, 324 (1936).

\bibitem{Braak2011} 
Braak, D. Integrability of the Rabi model. 
\textit{Phys. Rev. Lett.} {\bf 107}, 100401 (2011).

\bibitem{Bourassa2009} 
Bourassa, J., Gambetta,  J. M., Abdumalikov, A. A., Jr., Astafiev,  O., Nakamura, Y. \&  Blais, A. 
Ultrastrong coupling regime of cavity QED with phase-biased flux qubits.
\textit{Phys. Rev. B} {\bf 80}, 032109 (2009).

\bibitem{Niemczyk2010}
Niemczyk, T., Deppe, F., Huebl,  H., Menzel, E. P., Hocke, F., Schwarz,  M. J.,	Garc\'ia-Ripoll, J. J., Zueco, D., H\"ummer, T., Solano, E., Marx, A. \& Gross, R. 
Circuit quantum electrodynamics in the ultrastrong-coupling regime.
\textit{Nat. Phys.} {\bf 6}, 772 (2010).

\bibitem{Fedorov2010}
Fedorov,  A., Feofanov, A. K., Macha, P., Forn-D\'iaz, P., Harmans, C. J. P. M. \& Mooij, J. E.
Strong Coupling of a Quantum Oscillator to a Flux Qubit at Its Symmetry Point.
\textit{Phys. Rev. Lett.} {\bf 105}, 060503 (2010).

\bibitem{Diaz2010}
P. Forn-D\'iaz,  A., Lisenfeld, J., Marcos, D., Garc\'ia-Ripoll, J. J., Solano, E., Harmans, C. J. P. M., \& Mooij, J. E.
Observation of the Bloch-Siegert Shift in a Qubit-Oscillator System in the Ultrastrong Coupling Regime.
\textit{Phys. Rev. Lett.} {\bf 105}, 237001 (2010).

\bibitem{Anappara2009} 
Anappara, A. A., De Liberato, S., Tredicucci, A., Ciuti, C., Biasiol, G., Sorba, L. \&  Beltram, F. 
Signatures of the ultrastrong light-matter coupling regime.
\textit{Phys. Rev. B} {\bf 79}, 201303(R) (2009).

\bibitem{Gunter2009}
G\"unter, G., Anappara, A.A., Hees, J., Sell, A., Biasiol, G., Sorba, L., De Liberato, S., Ciuti, C., Tredicucci, A., Leitenstorfer, A., Huber, R.
Sub-cycle switch-on of ultrastrong light-matter interaction.
\textit{Nature} {\bf 458}, 178 (2009).

\bibitem{Todorov2010}
Todorov, Y., Andrews, A. M., Colombelli, R., De Liberato, S., Ciuti, C., Klang, P., Strasser, G. \&  Sirtori, C. 
Ultrastrong Light-Matter Coupling Regime with Polariton Dots.
\textit{Phys. Rev. Lett.} {\bf 105}, 196402 (2010).

\bibitem{Gustafsson2014} 
Gustafsson, M. V., Aref, T., Kockum, A. F., Ekstr\"om, M. K., Johansson, G., \& Delsing, P. 
Propagating phonons coupled to an artificial atom.
\textit{Science} {\bf 346}, 207 (2014).

\bibitem{Ciuti2005} 
Ciuti, C., Bastard, G. \& Carusotto, I.
Quantum vacuum properties of the intersubband cavity polariton field.
\textit{Phys. Rev. B} {\bf 72}, 115303 (2005).

\bibitem{Nataf2011} Nataf, P. \& Ciuti, C. Protected Quantum Computation with Multiple Resonators in Ultrastrong Coupling Circuit QED. \textit{Phys. Rev. Lett.} \textbf{107,} 190402 (2011). 

\bibitem{Emary2004} 
Emary, C. \& Brandes, T.
Phase transitions in generalized spin-boson (Dicke) models.
\textit{Phys. Rev. A} {\bf 69}, 053804 (2004).

\bibitem{Ciuti2006} 
Ciuti, C. \& Carusotto, I.
Input-output theory of cavities in the ultrastrong coupling regime: The case of time-independent cavity parameters.
\textit{Phys. Rev. A} {\bf 74}, 033811 (2006).

\bibitem{Deliberato2009} 
De Liberato, S., Gerace, D., Carusotto, I., \& Ciuti, C.
Extracavity quantum vacuum radiation from a single qubit.
\textit{Phys. Rev. A} {\bf 80}, 053810 (2009).

\bibitem{Meaney2010} 
Meaney, C. P., Duty, T., McKenzie, R. H., \& Milburn, G. J.
Jahn-Teller instability in dissipative quantum systems.
\textit{Phys. Rev. A} {\bf 81}, 043805 (2010).

\bibitem{Ashhab2010} 
Ashhab, S. \& Nori, F.
Qubit-oscillator systems in the ultrastrong-coupling regime and their potential for preparing nonclassical states.
\textit{Phys. Rev. A} {\bf 81}, 042311 (2010).

\bibitem{Ridolfo2012} 
Ridolfo, A., Leib, M., Savasta, S. \& Hartmann, M. J. Photon Blockade in the Ultrastrong Coupling Regime.
 \textit{Phys. Rev. Lett.} \textbf{109,} 193602 (2012).

\bibitem{Felicetti2014} 
Felicetti, S., Romero, G., Rossini, D., Fazio, R. \&  Solano, E.
Photon transfer in ultrastrongly coupled three-cavity arrays.
\textit{Phys. Rev. A} {\bf 89}, 013853 (2014).

\bibitem{Romero2012} Romero, G., Ballester, D., Wang, Y. M., Scarani, V., \& Solano, E. Ultrafast Quantum Gates in Circuit QED. \textit{Phys. Rev. Lett.} {\bf 108,} 120501 (2012).

\bibitem{Kyaw2014}
Kyaw, T. H., Felicetti, S., Romero, G., Solano, E. \&  Kwek, L. C.
Scalable quantum random-access memory with superconducting circuits.
\textit{Sci. Rep.} {\bf 5}, 8621 (2015); Z2 quantum memory implemented on circuit quantum electrodynamics. \textit{Proc. SPIE 9225, Quantum Communications and Quantum Imaging XII}, 92250B (October 8, 2014).

\bibitem{Shalibo2013} 
Shalibo, Y.,  Resh, R., Fogel, O., Shwa, D., Bialczak, R., Martinis, J. M. \&  Katz, N.
Direct Wigner Tomography of a Superconducting Anharmonic Oscillator.
\textit{Phys. Rev. Lett.} {\bf 110}, 100404 (2013).

\bibitem{Bertet2002} 
 Bertet, P., Auffeves,  A., Maioli, P.,  Osnaghi, S.,  Meunier, T.,  Brune, M.,  Raimond, J.-M., \&  Haroche S.
Direct Measurement of the Wigner Function of a One-Photon Fock State in a Cavity.
\textit{Phys. Rev. Lett.} {\bf 89}, 200402 (2002).

\bibitem{Deleglise2008}
Deleglise, S.,  Dotsenko, I.,  Sayrin, C.,  Bernu,  J.,  Brune, M.,  Raimond, J.-M. \&  Haroche, S.
Reconstruction of non-classical cavity field states with snapshots of their decoherence.
 \textit{Nature} {\bf 455}, 510 (2008).

\bibitem{JCmodel}
Jaynes, E. T.  \&  Cummings, F. W.
Comparison of quantum and semiclassical radiation theories with application to the beam maser.
 \textit{Proc. IEEE} {\bf 51}, 89 (1963).

\bibitem{Irish2007} 
Irish, E. K. 
Generalized Rotating-Wave Approximation for Arbitrarily Large Coupling
 \textit{Phys. Rev. Lett.} {\bf 99}, 173601 (2007).

\bibitem{Casanova2010} Casanova, J., Romero, G., Lizuain, I., Garc\'ia-Ripoll, J. J., \& Solano, E. Deep Strong Coupling Regime of the Jaynes-Cummings Model. \textit{Phys. Rev. Lett.} \textbf{105,} 263603 (2010).

\bibitem{DeLiberato2014}
De Liberato, S.
Light-Matter Decoupling in the Deep Strong Coupling Regime: The Breakdown of the Purcell Effect.
\textit{Phys. Rev. Lett.} {\bf 112}, 016401 (2014).

\bibitem{Koch2007} 
Koch, J.,  Yu, T. M., Gambetta, J., Houck,  A. A.,   Schuster, D. I.,  Majer, J.,  Blais, A.,  Devoret, M. H.,  Girvin, S. M. \&  Schoelkopf, R. J.
Charge-insensitive qubit design derived from the Cooper pair box.
\textit{Phys. Rev. Lett.} {\bf 76}, 042319 (2007).

\bibitem{Srinivasan2011} 
Srinivasan, S. J.,  Hoffman,  A. J.,  Gambetta, J. M. \&  Houck, A. A.
Tunable Coupling in Circuit Quantum Electrodynamics Using a Superconducting Charge Qubit with a V-Shaped Energy Level Diagram.
\textit{Phys. Rev. Lett.} {\bf 106}, 083601 (2011).

\bibitem{OpenQBook}  Breuer, H.-P. \& Petruccione, F., The Theory of Open Quantum Systems.
 \textit{Clarence Press}, Oxford, 2006.

\bibitem{Gambetta2007} 
Gambetta, J. M.,  Braff, W. A., Wallraff, A., Girvin, S. M. \&  Schoelkopf, R. J. 
Protocols for optimal readout of qubits using a continuous quantum nondemolition measurement.
\textit{Phys. Rev. A} {\bf 76}, 012325 (2007).

\bibitem{Lad2007}
Lupascu, A.,  Saito, S., Picot, T., de Groot, P. C., Harmans, C. J. P. M. \&  J. E. Mooij, J. E.
Quantum non-demolition measurement of a superconducting two-level system.
 \textit{Nat. Phys.} {\bf 3}, 119 (2007).

\bibitem{Steffen2006} 
Steffen, M., Ansmann, M.,   McDermott, R.,  Katz, N.,  Bialczak, R. C., Lucero, E.,  Neeley, M.,  Weig, E. M.,  Cleland, A. N. \&  Martinis, J. M. 
State Tomography of Capacitively Shunted Phase Qubits with High Fidelity.
\textit{Phys. Rev. Lett.} {\bf 97}, 050502 (2006).


\end{thebibliography}

\begin{thebibliography}{9}

\bibitem{Braak2011_sup} 
D.~Braak, Phys. Rev. Lett. {\bf 107}, 100401 (2011).

\bibitem{Srinivasan2011_sup} 
S. J. Srinivasan, A. J. Hoffman, J. M. Gambetta, and A. A. Houck, Phys. Rev. Lett. {\bf 106}, 083601 (2011).

\end{thebibliography}
\end{document}